\documentclass[draft,showpacs,showkeys,eqsecnum,nofootinbib,aps]{revtex4}
\renewcommand{\theequation}{\arabic{equation}}
\def\beq{\begin{equation}}
\def\eeq{\end{equation}}
\def\bea{\begin{eqnarray}}
\def\eea{\end{eqnarray}}
\def\nn{\nonumber}

\begin{document}
\title{Complete higher dimensional global embedding structures of various black holes}
\author{Soon-Tae Hong}
\email{soonhong@ewha.ac.kr} \affiliation{Department of Science
Education, Ewha Womans University, Seoul 120-750 Korea}
\date{November 29, 2003}
\begin{abstract}
We study global flat embeddings inside and outside of event horizons of black holes 
such as Schwarzschild and Reissner-Nordstr\"{o}m black holes, and of de Sitter space.  
On these overall patches of the curved manifolds we investigate four accelerations 
and Hawking temperatures by introducing relevant Killing vectors.
\end{abstract}
\maketitle

\section{Introduction}
\setcounter{equation}{0}
\renewcommand{\theequation}{\arabic{section}.\arabic{equation}}

Ever since the discovery that thermodynamic properties of black holes in anti-de
Sitter (AdS) spacetime are dual to those of a field theory in one dimension
fewer~\cite{rnads}, there has been of much interest in the Reissner-Nordstr\"om (RN) black
hole~\cite{rn}, which now becomes a prototype example for studying this AdS/CFT
correspondence~\cite{wit}.  It is also well understood that, in differential
geometry four dimensional Schwarzschild metric~\cite{sch} is not embedded in 
$R^{5}$~\cite{spivak75}.  Moreover, $n$ dimensional spacetime has been shown to be embedded 
into $d$ dimensional pseudo-Euclidean space with dimensionality 
$n\leq d \leq n(n+1)/2$~\cite{eisen49}, so that more than ten dimensions cannot be 
required to embed any four dimensional solution of Einstein equations with arbitrary 
energy-momentum tensor.  

Recently, (5+1) dimensional global embedding Minkowski space (GEMS) structure for the region 
outside the event horizon of the Schwarzschild black hole has been 
obtained~\cite{deser97,deser99} to investigate a thermal Hawking
effect on a curved manifold~\cite{hawk75} associated with an Unruh effect~\cite{unr} in
these higher dimensional space time where the usual black hole detectors are mapped into
Rindler observers with the correct temperatures as determined from their constant
accelerations.  The multiply warped product manifold associated
with the (3+1) RN metric has been also studied to investigate the geometrical
properties ``inside" the event horizons~\cite{hong02math}.  In this analysis, all the 
expressions of the Ricci components and the Einstein scalar curvature were shown to 
be form invariant both in the exterior
and interior of the outer event horizon without discontinuities.  It has been also shown 
in the GEMS approach to the (2+1) dimensional black holes that
the uncharged and charged Banados-Teitelboim-Zanelli (BTZ) black holes~\cite{btz,cal}
are embedded in (2+2)~\cite{deser97,hong00prd} and (3+2) dimensions~\cite{hongplb}, while the uncharged
and charged black strings are embedded in (3+1) and (3+2) dimensions~\cite{hong01prd},
respectively.  Note that the dual solutions of the BTZ black holes are related to the solutions in
the string theory, so-called (2+1) black strings~\cite{horowitz93,callan85}.  Moreover, in the
warped product approach to the BTZ black hole, all the Ricci components and the Einstein scalar
curvature have the form invariant expressions both inside and outside the outer event horizon without
discontinuities~\cite{honggrg}.  Quite recently, the BTZ black holes
were further analyzed to yield the complete GEMS structures ``outside and
inside" event horizons~\cite{hongplb}.

On the other hand, exploiting the Kruskal extension of the
Schwarzschild black hole, the coordinate singularity at the event
horizon $r=2m$ can be eliminated to yield in (3+1) dimensional
spacetime~\cite{kru} \beq
ds^2=\frac{32m^{3}e^{-r/2m}}{r}(dT^2-dX^{2})-r^{2}(d\theta^{2}+\sin^{2}\theta
d\phi^{2}), \label{krumetric} \eeq via coordinate transformations
\beq
T=\frac{1}{2}\left(e^{(t+r_{*})/4m}-e^{-(t-r_{*})/4m}\right),~~~
X=\frac{1}{2}\left(e^{(t+r_{*})/4m}+e^{-(t-r_{*})/4m}\right),
\label{krucoor} \eeq with the Regge-Wheeler tortoise coordinate
$r_{*}$ defined by \beq r_{*}=r+2m\ln\left(\frac{r}{2m}-1\right).
\eeq Note that the metric (\ref{krumetric}) is not a flat metric
and the Kruskal coordinates (\ref{krucoor}) are inconvenient for
analyzing the asymptotically flat region
$r\rightarrow\infty$~\cite{wald}.  However, the higher dimensional
flat embedding of the Schwarzschild black hole, which will be
discussed later, is well-defined in the asymptotically flat
region~\cite{deser99}.  Even though the region outside the event
horizon of the Schwarzschild black hole has been nicely described
in this GEMS structure to investigate the thermodynamic properties
of black hole in terms of the Unruh effect, the region inside the
event horizon remains intact with a brief comment that the
extension to the interior region is just the maximal Kruskal
one~\cite{deser99}.

In this paper we will investigate the region ``inside" the event horizon of the Schwarzschild
black hole to construct explicitly complete embedding solutions, four accelerations and
Hawking temperatures inside the event horizons.  Moreover, we will construct embedding 
solutions inside and outside the event horizons of the de Sitter (dS) space and RN black 
hole to investigate their four accelerations and Hawking temperatures on the overall patches of 
these curved manifolds.  We will consider the Schwarzschild black hole in section 2, the 
RN black hole in section 3 and the dS spaces in section 4,
respectively.

\section{Schwarzschild black hole}
\setcounter{equation}{0}
\renewcommand{\theequation}{\arabic{section}.\arabic{equation}}


We begin with a brief recapitulation of the results of the global embedding Minkowski
space (GEMS) approach~\cite{deser97}, for the (3+1) dimensional Schwarzschild 
black hole~\cite{sch} whose four-metric is given by
\beq
ds^2=N^2dt^2-N^{-2}dr^{2}-r^{2}(d\theta^{2}+\sin^{2}\theta d\phi^{2}),
\label{fourmetric}
\eeq
where the exterior lapse function is
\beq
N^{2}=1-\frac{2m}{r}= \frac{r- r_{H}}{r},
\label{schlapse}
\eeq
with the event horizon $r_{H}=2m$.  The (5+1) minimal Schwarzschild GEMS
\beq
ds^{2}=\eta_{MN}dz^{M}dz^{N},~~~\eta_{MN}={\rm diag}(+-----)
\label{ds2eta}
\eeq
is then given by the coordinate
transformations for $r\geq r_{H}$ as
follows~\cite{deser97}
\bea
z^{0}&=&k_{H}^{-1}\left(\frac{r-r_{H}}{r}\right)^{1/2}\sinh k_{H}t,
\nonumber \\
z^{1}&=&k_{H}^{-1}\left(\frac{r-r_{H}}{r}\right)^{1/2}\cosh k_{H}t,
\nonumber \\
z^{2}&=&\int dr \left(\frac{r_{H}(r^{2}+r_{H}r+r_{H}^{2})}{r^{3}}\right)^{1/2}
\equiv f(r,r_{H}),
\nonumber \\
z^{3}&=&r\sin\theta\cos\phi,
\nonumber\\
z^{4}&=&r\sin\theta\sin\phi,
\nonumber\\
z^{5}&=&r\cos\theta
\label{sch2}
\eea
with the surface gravity $k_{H}=1/2r_{H}$.

Now, to construct the GEMS inside the event horizon $r\leq r_{H}$,
we use the four-metric \beq
ds^2=\bar{N}^{-2}dr^{2}-\bar{N}^2dt^2-r^{2}(d\theta^{2}+\sin^{2}\theta
d\phi^{2}), \label{fourmetric2} 
\eeq 
where the interior lapse function is then given by 
\beq 
\bar{N}^{2}=-1+\frac{2m}{r}=
\frac{r_{H}-r}{r}. \label{schlapse1} 
\eeq 
We introduce an ansatz for the coordinate transformations for $r\leq r_{H}$ 
\beq
z^{M}=(r\cosh k_{H}t\sin R,r\sinh k_{H}t\sin
R,f,r\sin\theta\cos\phi,r\sin\theta\sin\phi, r\cos\theta)
\label{zm3} \eeq where $f$ is defined as
in (\ref{sch2}) and $\sin R$ will be fixed later, to obtain the
(5+1) GEMS structure (\ref{ds2eta}) for $r\leq r_{H}$ which yields
\beq 
ds^{2}=-\bar{N}^{2}dt^{2}+(\sin Rdr+r\cos R dR)^{2}-dr^{2}
-r^{2}(d{\theta}^{2} + \sin^{2}\theta d{\phi}^{2})-df^{2}.
\label{ds2mschm} 
\eeq 
With the ansatz for $\sin R$ and $\cos R$ 
\beq 
\sin R=\frac{1}{k_{H}r} \left(\frac{r_{H}-r}{r}\right)^{1/2},~~~
\cos R=\frac{1}{k_{H}r} \left(\frac{k_{H}^{2}r^{3}-r_{H}+r}{r}\right)^{1/2},\label{ansatz} \eeq 
(\ref{ds2mschm}) reproduce the Schwarzschild
four-metric (\ref{fourmetric2}) associated with the interior lapse function (\ref{schlapse1}), 
to arrive at the (5+1) GEMS structure (\ref{ds2eta}) with the coordinate transformations 
in the region $r\leq r_{H}$ 
\bea
z^{0}&=&k_{H}^{-1}\left(\frac{r_{H}-r}{r}\right)^{1/2}\cosh
k_{H}t,
\nonumber \\
z^{1}&=&k_{H}^{-1}\left(\frac{r_{H}-r}{r}\right)^{1/2}\sinh k_{H}t,
\nonumber \\
z^{2}&=&f(r,r_{H}),
\label{sch52o}
\eea
with $(z^{3},z^{4},z^{5})$ in (\ref{sch2}).  Here note that the coordinate singularity at $r=r_{H}$ does not
appear in the transformations (\ref{sch2}) and (\ref{sch52o}).  Next, introducing the Killing
vector $\xi=\partial_{r}$ inside the event horizon we obtain the four acceleration
\beq
a_{4}=\frac{r_{H}}{2r^{3/2}(r_{H}-r)^{1/2}}
\eeq
and the local Hawking temperature inside the event horizon
\beq
T=\frac{a_{6}}{2\pi}=\frac{1}{4\pi r_{H}}\left(\frac{r}{r_{H}-r}\right)^{1/2}.
\eeq
Note that the role of timelike Killing vector $\xi=\partial_{t}$ defined outside the event horizon is
replaced by that of the ``timelike" Killing vector $\xi=\partial_{r}$ in the ``interior" Schwarzschild
solution associated with the four-metric (\ref{fourmetric2}).  The black hole temperature is then given by
\beq
T_{0}=\frac{1}{4\pi r_{H}}.
\label{scht02}
\eeq


Next, in order to investigate further the global embedding structure outside the event horizon,
we can take the ansatz for $z^{M}$
\beq
z^{M}=(r\sinh k_{H}t\sinh R,r\cosh k_{H}t\sinh R,f,r\sin\theta\cos\phi,r\sin\theta\sin\phi, r\cos\theta)
\label{zmsch1}
\eeq
with $f$ in (\ref{sch2}) and $\sinh R$ and $\cosh R$ defined as
\beq
\sinh R=\frac{1}{k_{H}r} \left(\frac{r-r_{H}}{r}\right)^{1/2},~~~
\cosh R=\frac{1}{k_{H}r} \left(\frac{k_{H}^{2}r^{3}+r-r_{H}}{r}\right)^{1/2},\label{ansatz1sch}
\eeq
to reproduce the Schwrzschild metric (\ref{fourmetric}) associated with the lapse function 
(\ref{schlapse}) so that we can reconstruct the (5+1) GEMS structure (\ref{ds2eta}) with the 
coordinate transformations (\ref{sch2}) for $r\geq r_{H}$.  Moreover, introducing the
Killing vector $\xi=\partial_{t}$ outside the event horizon we obtain the four acceleration
\beq
a_{4}=\frac{r_{H}}{2r^{3/2}(r-r_{H})^{1/2}},
\label{a4schw}
\eeq
and the local Hawking temperature~\cite{deser97}
\beq
T=\frac{1}{4\pi r_{H}}\left(\frac{r}{r-r_{H}}\right)^{1/2}.
\label{scht2}
\eeq
Moreover, the black hole temperature $T_{0}$ is the same as that inside
the event horizon in (\ref{scht02}).

\section{Reissner-Nordstr\"{o}m black hole}
\setcounter{equation}{0}
\renewcommand{\theequation}{\arabic{section}.\arabic{equation}}


Now, in order to investigate the GEMS structure in the range between
the inner and outer event horizons for the nonextremal case of the (3+1)
dimensional RN black hole~\cite{rn}, we introduce
the four-metric (\ref{fourmetric2}) with the lapse function
\beq
\bar{N}^{2}=-1+\frac{2m}{r}-\frac{Q^{2}}{r^{2}}.
\label{schlapse2}
\eeq
Note that two event horizons $r_{\pm}(Q)$ satisfy the equations $0=-1+2m/r_{\pm}-Q^{2}/r_{\pm}^{2}$,
and the lapse function can be rewritten in terms of these outer and inner horizons
\begin{equation}
\bar{N}^{2}=\frac{(r_{+}-r)(r-r_{-})}{r^{2}}
\label{n20002}
\end{equation}
which is well defined for $r_{-}\leq r\leq r_{+}$, and the parameters $m$
and $Q$ can be rewritten in terms of $r_{\pm}$ as follows
\beq
m=\frac{r_{+}+r_{-}}{2},~~~Q^{2}=r_{+}r_{-}.
\label{qandl}
\eeq

Introducing in the region $r_{-}\leq r\leq r_{+}$
\beq
z^{M}=(r\cosh k_{H}t\sin R,r\sinh k_{H}t\sin R,f,r\sin\theta\cos\phi,r\sin\theta\sin\phi, r\cos\theta, g)
\label{rnxm32}
\eeq
with $f$ and $g$ fixed later and the surface gravity $k_{H}=(r_{+}-r_{-})/2r_{+}^{2}$, we obtain
\beq
ds^{2}=-\bar{N}^{2}dt^{2}+(\sin Rdr+r\cos R dR)^{2}-dr^{2}-r^{2}(d{\theta}^{2} + \sin^{2}\theta d{\phi}^{2})
-df^{2}+dg^{2}.
\eeq
With the ansatz for $\sin R$ and $\cos R$
\beq
\sin R=\frac{1}{k_{H}r} \left(\frac{(r_{+}-r)(r-r_{-})}{r^{2}}\right)^{1/2},~~~
\cos R=\frac{1}{k_{H}r} \left(\frac{k_{H}^{2}r^{4}-(r_{+}-r)(r-r_{-})}{r^{2}}\right)^{1/2},
\label{rnansatz2}
\eeq
we can construct, after some algebra, the (5+2) minimal RN
GEMS structure
\beq
ds^{2}=\eta_{MN}dz^{M}dz^{N},~~~\eta_{MN}={\rm diag}(+-----+)
\label{ds2mn2}
\eeq
with the coordinate transformations for $r_{-}\leq r\leq r_{+}$
\bea
z^{0}&=&k_{H}^{-1}\left(\frac{(r_{+}-r)(r-r_{-})}{r^{2}}\right)^{1/2}\cosh k_{H}t,
\nonumber \\
z^{1}&=&k_{H}^{-1}\left(\frac{(r_{+}-r)(r-r_{-})}{r^{2}}\right)^{1/2}\sinh k_{H}t,
\nonumber \\
z^{2}&=&\int dr \left(\frac{r^{2}(r_{+}+r_{-})+r_{+}^{2}(r+r_{+})}{r^{2}(r-r_{-})}\right)^{1/2}
\equiv f(r,r_{+},r_{-}),
\nonumber \\
z^{3}&=&r\sin\theta\cos\phi,
\nonumber\\
z^{4}&=&r\sin\theta\sin\phi,
\nonumber\\
z^{5}&=&r\cos\theta,\nn\\
z^{6}&=&\int dr \left(\frac{4r_{+}^{5}r_{-}}{r^{4}(r_{+}-r_{-})^{2}}\right)^{1/2}
\equiv g(r,r_{+},r_{-}).
\label{rn2}
\eea
Exploiting the Killing vector $\xi=\partial_{r}$ for $r_{-}\leq r\leq r_{+}$, as in the interior
Schwarzschild black hole solution case, we obtain the four acceleration
\beq
a_{4}=\frac{(r_{+}+r_{-})r-2r_{+}r_{-}}{2r^{2}[(r_{+}-r)(r-r_{-})]^{1/2}},
\eeq
the local Hawking temperature
\beq
T=\frac{a_{7}}{2\pi}=\frac{(r_{+}-r_{-})r}{4\pi r_{+}^{2}\left[(r_{+}-r)(r-r_{-})\right]^{1/2}},
\eeq
and the black hole temperature
\beq
T_{0}=\frac{r_{+}-r_{-}}{4\pi r_{+}^{2}}.\label{rnt0}
\eeq


Next, for the case of the range inside the inner event horizon $r\leq r_{-}$, we introduce
$z^{M}$
\beq
z^{M}=(r\sinh k_{H}t\sinh R,r\cosh k_{H}t\sinh R,f,r\sin\theta\cos\phi,r\sin\theta\sin\phi, r\cos\theta,g)
\label{zmrn1}
\eeq
with $f$ and $g$ in (\ref{rn2}), to arrive at the GEMS
structure (\ref{ds2mn2}) yielding
\beq
ds^{2}=N^{2}dt^{2}-(\sinh Rdr+r\cosh R dR)^{2}-dr^{2}
-r^{2}(d{\theta}^{2} + \sin^{2}\theta d{\phi}^{2})-df^{2}+dg^{2},
\label{ds2mrnp}
\eeq
where the RN lapse function for $r\leq r_{-}$ is given by
\beq
N^{2}=1-\frac{2m}{r}+\frac{Q^{2}}{r^{2}}=\frac{(r_{+}-r)(r_{-}-r)}{r^{2}}.
\label{rnlapse3}
\eeq
With the ansatz for $\sinh R$ and $\cosh R$
\beq
\sinh R=\frac{1}{k_{H}r} \left(\frac{(r_{+}-r)(r_{-}-r)}{r^{2}}\right)^{1/2},~~~
\cosh R=\frac{1}{k_{H}r} \left(\frac{k_{H}^{2}r^{4}+(r_{+}-r)(r_{-}-r)}{r^{2}}\right)^{1/2},
\label{rnansatz3}
\eeq
we can produce the four-metric (\ref{fourmetric}) associated with the RN lapse function 
(\ref{rnlapse3}), and the coordinate transformations of the (5+2) GEMS structure (\ref{ds2mn2}) 
for $r\leq r_{-}$ is given by
\bea
z^{0}&=&k_{H}^{-1}\left(\frac{(r_{+}-r)(r_{-}-r)}{r^{2}}\right)^{1/2}\sinh k_{H}t,
\nonumber\\
z^{1}&=&k_{H}^{-1}\left(\frac{(r_{+}-r)(r_{-}-r)}{r^{2}}\right)^{1/2}\cosh k_{H}t,
\nonumber\\
z^{2}&=&f(r,r_{+},r_{-}),
\nonumber\\
z^{6}&=&g(r,r_{+},r_{-}),
\label{rn23}
\eea
with $(z^{3},z^{4},z^{5})$ and $f$ and $g$ in (\ref{rn2}).  Note that the coordinate 
singularities at $r=r_{\pm}$ do not appear in the transformation (\ref{rn23}) as 
well as in (\ref{rn2}).  Introducing the Killing vector $\xi=\partial_{t}$ in the region 
$r\leq r_{-}$ we obtain the four acceleration
\beq
a_{4}=\frac{(r_{+}+r_{-})r-2r_{+}r_{-}}{2r^{2}[(r_{+}-r)(r_{-}-r)]^{1/2}}
\eeq
and the local Hawking temperature
\beq
T=\frac{(r_{+}-r_{-})r}{4\pi r_{+}^{2}\left[(r_{+}-r)(r_{-}-r)\right]^{1/2}},
\eeq
and the black hole temperature $T_{0}$ is the same as (\ref{rnt0}) of the
region $r_{-}\leq r\leq r_{+}$.


Finally, it seems appropriate to comment on the GEMS structure outside the outer event
horizon $r\geq r_{+}$ of the RN black hole.  In this region we have the four-metric
(\ref{fourmetric}) with the lapse function which is the same as (\ref{rnlapse3}) of
the region $r\leq r_{-}$.  Moreover, after some algebra we can obtain the (5+2) GEMS
structure (\ref{ds2mn2}) for $r\geq r_{+}$ with the same coordinate transformations 
(\ref{rn23})~\cite{deser97,kps99}.  Introducing the Killing vector $\xi=\partial_{t}$ again 
outside the outer event horizon, we can obtain the four acceleration
\beq
a_{4}=\frac{(r_{+}+r_{-})r-2r_{+}r_{-}}{2r^{2}[(r-r_{+})(r-r_{-})]^{1/2}},
\eeq
the local Hawking temperature~\cite{deser97}
\beq
T=\frac{(r_{+}-r_{-})r}{4\pi r_{+}^{2}\left[(r-r_{+})(r-r_{-})\right]^{1/2}},
\label{rnhawking}
\eeq
and the black hole temperature $T_{0}$ is the same as (\ref{rnt0}).  Note that $a_{4}$, $T$
and $T_{0}$ in this region are the same as those in the region $r\leq r_{-}$.  Moreover, in the 
limit of $r_{-}\rightarrow0$, these quantities  $a_{4}$, $T$ and $T_{0}$ are reduced to those 
of the Schwarzschild black hole solution, (\ref{a4schw}), (\ref{scht2}) and (\ref{scht02}).

\section{ de Sitter space}
\setcounter{equation}{0}
\renewcommand{\theequation}{\arabic{section}.\arabic{equation}}


In this section, we begin with the (3+1) dimensional dS space
described by the four-metric (\ref{fourmetric2}) with the exterior
lapse function \beq
\bar{N}^{2}=-1+\frac{r^2}{l^2}=\frac{r^{2}-r_{H}^{2}}{r_{H}^2},
\label{dslapse} \eeq with the event horizon $r_{H}=l$ satisfying
the equation $0=-1+r_{H}^2/l^2$. In order to construct the GEMS
outside the event horizon, we introduce 
\beq 
z^{M}=(\cosh k_{H}t\sinh R,\sinh k_{H}t\sinh R,r\sin\theta\cos\phi,r\sin\theta\sin\phi, r\cos\theta)
\label{dszm3} \eeq  
with the surface gravity $k_{H}=1/r_{H}$, to obtain for $r\geq r_{H}$ 
\beq ds^{2}=-\bar{N}^{2}dt^{2}+\cosh^{2}R dR^{2}-dr^{2}
-r^{2}(d{\theta}^{2} + \sin^{2}\theta d{\phi}^{2}). \label{dsds22}
\eeq 
With the ansatz for $\sinh R$ and $\cosh R$, 
\beq 
\sinh R=\left(r^{2}-r_{H}^{2}\right)^{1/2},~~~ 
\cosh R=\left(1+r^{2}-r_{H}^{2}\right)^{1/2},
\label{dsansatz2} \eeq
the four-metric (\ref{dsds22}) yields the (4+1) GEMS structure for $r\geq r_{H}$
\beq ds^{2}=\eta_{MN}dz^{M}dz^{N},~~~\eta_{MN}={\rm diag}(+----),
\label{ds2mnds} \eeq with the coordinate transformations \bea
z^{0}&=&k_{H}^{-1}\left(\frac{r^2-r_{H}^2}{r_{H}^2}\right)^{1/2}
        \cosh k_{H}t, \nonumber \\
z^{1}&=&k_{H}^{-1}\left(\frac{r^2-r_{H}^2}{r_{H}^2}\right)^{1/2}
        \sinh k_{H}t,\nn\\
z^{2}&=&r\sin\theta\cos\phi,
\nonumber\\
z^{3}&=&r\sin\theta\sin\phi,
\nonumber\\
z^{4}&=&r\cos\theta.
\label{ds22}
\eea
Introducing the Killing vector $\xi=\partial_{r}$ outside the event horizon as in the 
previous sections, we obtain the four acceleration
\beq
a_{4}=\frac{r}{r_{H}(r^{2}-r_{H}^{2})^{1/2}},\eeq
the local Hawking temperature
\beq
T=\frac{a_{5}}{2\pi}=\frac{1}{2\pi \left(r^{2}-r_{H}^{2}\right)^{1/2}},
\label{dst2}
\eeq
and the temperature $T_{0}$ given by
\beq
T_{0}=\frac{1}{2\pi r_{H}}.
\label{dstzero1}
\eeq


Next, for the case of the range inside the event horizon $r\leq
r_{H}$ where we have the four-metric (\ref{fourmetric}) with the interior lapse function 
\beq
N^{2}=1-\frac{r^2}{l^2}=\frac{r_{H}^{2}-r^{2}}{r_{H}^2},
\eeq
we introduce $z^{M}$ 
\beq 
z^{M}=(\sinh k_{H}t\sin R,\cosh k_{H}t\sin
R,r\sin\theta\cos\phi,r\sin\theta\sin\phi, r\cos\theta)
\label{zmsch} \eeq  
to yield 
\beq ds^{2}=N^{2}dt^{2}-\cos^{2}R dR^{2}-dr^{2}
-r^{2}(d{\theta}^{2} + \sin^{2}\theta d{\phi}^{2}). \label{dsds21}
\eeq Now we can readily check that (\ref{dsds21}) can be satisfied
with the ansatz for $\sin R$ and $\cos R$ 
\beq 
\sin R=\left(r_{H}^{2}-r^{2}\right)^{1/2},~~~
\cos R=\left(1-r_{H}^{2}+r^{2}\right)^{1/2},
\label{dsansatz1} 
\eeq 
to produce the (4+1) GEMS structure (\ref{ds2mnds}) for $r\leq r_{H}$
associated with the coordinate transformations~\cite{deser97} \bea
z^{0}&=&k_{H}^{-1}\left(\frac{r_{H}^2-r^2}{r_{H}^2}\right)^{1/2}
        \sinh k_{H}t, \nonumber \\
z^{1}&=&k_{H}^{-1}\left(\frac{r_{H}^2-r^2}{r_{H}^2}\right)^{1/2}
        \cosh k_{H}t, \nonumber \\
\label{ds21}
\eea
with $(z^{2},z^{3},z^{4})$ in (\ref{ds22}).  Note that the coordinate singularity 
at $r=r_{H}$ does not appear any more in the
transformation (\ref{ds21}) and (\ref{ds22}), in which we can readily find the 
identity $\eta_{MN}z^{M}z^{N}=-l^{2}$.  Introducing the Killing
vector $\xi=\partial_{t}$ we also obtain the four acceleration
\beq
a_{4}=\frac{r}{r_{H}(r_{H}^{2}-r^{2})^{1/2}}
\eeq
and the local Hawking temperature~\cite{deser97}
\beq
T=\frac{1}{2\pi (r_{H}^{2}-r^{2})^{1/2}}.
\label{dst1}
\eeq
Moreover, the temperature $T_{0}$ is the same as that outside the event horizon given in (\ref{dstzero1}).

\section{Conclusions}
\setcounter{equation}{0}
\renewcommand{\theequation}{\arabic{section}.\arabic{equation}}

In conclusion, we have constructed the complete embedding solutions, four accelerations and
Hawking temperatures inside and outside the event horizons of the dS space, Schwarzschild
and RN black holes to explicitly calculate four accelerations and Hawking temperatures
on the overall patches of these curved manifolds by introducing the relevant Killing vectors.
It was shown in these manifolds that the temperatures $T_{0}$ are identical on these
overall patches, while the four accelerations $a_{4}$ and local Hawking temperatures $T$
have different expressions dependent on the interiors and exteriors of the event horizons.

\acknowledgments
The author acknowledges financial support from the Korea Science and Engineering Foundation
Grant R01-2000-00015.


\end{document}